# STRUCTURAL AND ELECTRICAL PROPERTIES OF CDS THIN FILMS SPIN COATED ON GLASS SUBSTRATES

P. Samarasekara and P.A.S. Madushan

Department of Physics, University of Peradeniya, Peradeniya, Sri Lanka

*Abstract*

CdS samples were synthesized on amorphous glass and conductive glass substrates at room temperature using spin coating technique and subsequently annealed at five different temperatures in air for one hour. The speed and time of spin coating were varied initially to optimize the properties of the samples. FTIR spectroscopy was used in this experiment to analyze CdS bond formation variation with annealing temperature. A liquid junction photocell with $KI/I_2$ electrolyte was prepared to measure the photovoltaic properties. Photovoltage, photocurrent and corresponding output power were measured in order to determine the output power.

*Keywords:* CdS, FTIR, photocell, spin coating, thin films

## 1. Introduction:

The structure of Cadmium Sulfide (CdS) is solid hexagonal or cubic crystal. Thin films of CdS can be used as window material of Cds/CdTe solar cells and other electro optical devices due to its high band gap. CdS is a n-type yellowish semiconductor material with direct band gap 2.42 eV at room temperature. CdS films have been prepared using many different techniques such as screen printing [1], vacuum evaporation [2], electron beam evaporation method [3] and chemical bath deposition [4]. The refractive index of CdS nanoparticles is lower than its bulk state due to the quantum confinement effect, and CdS layers exhibit the wave guiding properties. Concentration and synthesis technique determine properties of CdS related nanoparticles and quantum dots. The quantum dots are considered as promising candidates for optoelectronic applications including light emitting diodes (LED). Several methods have been used to fabricate nanostructured materials in solar energy conversion. Quantum dots can be used as frequency converters to match the spectrum of the incoming radiation to the spectral efficiency of the solar cell. Highly crystalline and transparent CdS films have been deposited on a glass substrate by electron beam evaporation technique. The structural and optical properties of the films were investigated. The X-ray diffraction analysis revealed that the CdS films have a hexagonal structure and exhibit preferred orientation along the (002) plane. UV-visible spectra of CdS films indicate that the absorption edge becomes steeper, and the band gap present fluctuation changes in the range of 2.389–2.448 eV as the substrate temperature increased [3].

CdS indicates some magnetic properties [5]. Magnetic thin films also find potential applications in memory and microwave devices. Second and third order perturbed Heisenberg Hamiltonian was used to describe the magnetic properties of ferromagnetic and ferrite films by us [6-10]. Thin films of Lithium mixed ferrite, multi walled carbon nanotubes, $Cu_2O/CuO$ layers, Nickel ferrite and zinc oxide have been fabricated [11-15]. Electrical properties of $Cu_2O/CuO$ layers and ZnO films have been studied by us [13, 15]. A p-n junction could be prepared using chemically synthesized $Cu_2O/CuO$ layers [13]. Very high photovoltage could be obtained using reactive dc sputtered ZnO films in $KI/I_2$ electrolyte [15]. In this report, the structural and electrical properties of spin coated CdS films have been explained.

## 2. Experimental:

In the preparation of five chemical compounds for sol-gel, two chemical compound used as additive of solution and other three were used as Cd and S agents and solvent. In this process, CdS was embedded in polyethylene glycol based solution. A polyethylene glycol (PEG 400, Merck) sol was prepared by mixing 0.6 ml of PEG with 8.9 ml of ethanol and 0.5 ml of acetic acid while stirring for one hour. Cadmium hydroxide and thiourea were used as the Cd and S supply agents, respectively. These agents were dissolved in ethanol while stirring for one hour. Thereafter, this solution was slowly added to the PEG sol with vigorous stirring and was stirred for four hours to obtain the final solution for thin film deposition. The spin-coating technique was used to prepare thin films on amorphous glass and conductive glass substrates. The samples were prepared at rotation speeds of 900, 1200 and 2000 rpm. The films were subsequently annealed in air at 100, 200, 300 400 and 500 °C for one hour in order to remove the solvent and residual organics.

FTIR spectroscopy was used in this experiment to analyze CdS bond formation variation with annealing temperature. For FTIR spectroscopy, pellets were prepared using KBr and CdS with mass ratio 40:1 initially, a background analysis was performed using a pure KBr pellet, and then FTIR peaks were measured to study the variation of Cd-S bond formation at five different annealing temperatures given above.

For photoconductivity measurements, CdS thin films were spin-coated on Fluorine-doped Tin Oxide (FTO) coated glass substrates. Then these films were studied using a solar cell simulation system. For this process $KI/I_2$ solution was used as an electrolyte, this solution made by adding 1.006ml acetonitrile, 2.5132ml of ethylene carbonate, and 0.217g of KI and 0.12 g of $I_2$.

## 3. Results and discussion:

Figure 1 shows the FTIR absorption Spectroscopy annealed at 100 and 500 $^0$C for PEG 0.05ml, 0.5M CdS.

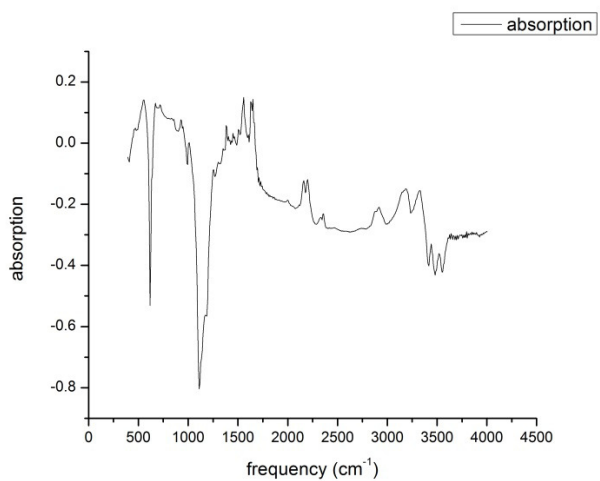

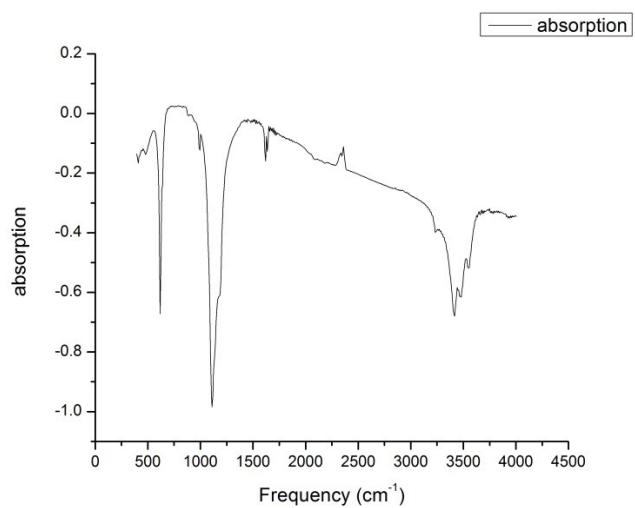

Fgiure 1: a) FTIR absorption Spectroscopy for (PEG 0.05ml, 0.5M CdS Annealed at 100 $^0$C)

b) FTIR absorption Spectroscopy for (PEG 0.05ml, 0.5M CdS Annealed at 500 $^0$C)

The FTIR spectrum of thin film samples shows different absorptions at peaks related to different chemical bonds. These peaks are listed in table 1.

| Wave number (cm$^{-1}$) | Chemical bond |
|---|---|
| 617.22 | Cd-S |
| 1118.71 | S-O (SO$_2$) |
| 1404.17 | C-C |
| 1635.63 | C-C |
| 3433.33 | O-H (H$_2$O) |

Table 1: The wave number and chemical bond related to absorption peak in the spectrum

Thereafter, the variation of Cd-S bond formation with annealing temperature was investigated by studying the wave number 617.22 cm$^{-1}$. This study indicates that the Cd-S bond formation also increases with the annealing temperature as given in figure 2.

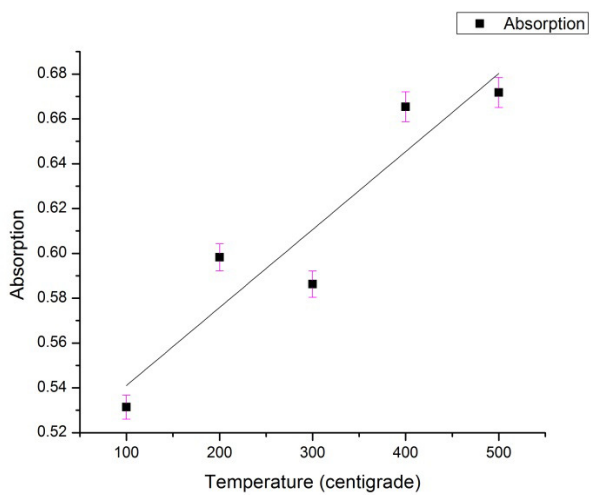

Figure 2: Variation of absorption due to Cd-S bond with annealing temperature.

The open circuit voltage and closed circuit current were measured to understand the variation of these two quantities with temperature using the solar simulator. The counter electrode was a platinum plate with area of 1.5x1.5cm$^2$. The area of CdS film was 1x1cm$^2$. A liquid junction photocell with CdS and KI/I$_2$ electrolyte was prepared for this. Because CdS is a n-type semiconductor, the KI/$I_2$ electrolyte acts as a p-type semiconductor. The redox couple in this photocell is $I_3^-/I^-$. Figure 3 shows the graphs of photocurrent versus photovoltage for liquid junction photocells prepared using CdS thin film samples annealed at 100 and 500 $^0$C.

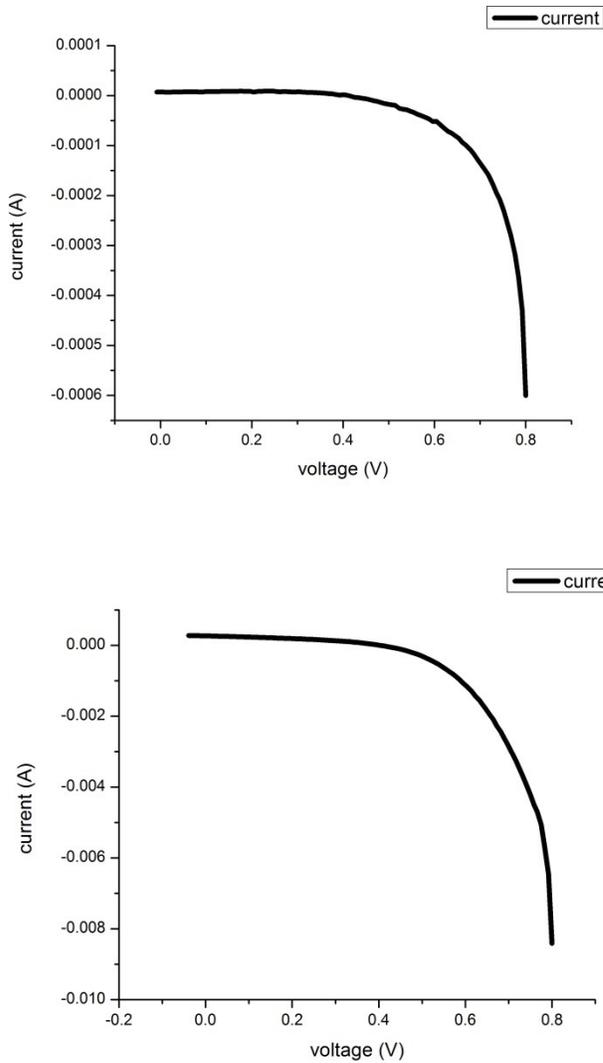

Figure 3: a) Photocurrent versus photo voltage for PEG 0.05ml, 0.5M CdS annealed at 100 $^0$C
b) Photocurrent versus photo voltage for PEG 0.05ml, 0.5M CdS annealed 500 $^0$C.

Table 2 shows the photovoltage, photocurrent and output power of liquid junction photocells prepared using CdS thin film samples annealed at different temperatures.

| Annealing temperature ($^0$C) | $V_{oc}$ (V) | $I_{sc}$ (A) | $I_{sc}*V_{oc}$ (VA) |
|---|---|---|---|
| 200 | 0.1987 | $1.9100*10^{-6}$ | $6.78*10^{-7}$ |
| 300 | 0.3194 | $5.2596*10^{-6}$ | $1.766*10^{-6}$ |
| 400 | 0.6357 | $6.5275*10^{-5}$ | $3.8827*10^{-5}$ |
| 500 | 0.4023 | $2.6367*10^{-4}$ | $1.067*10^{-4}$ |

Table 2: $I_{sc}$ and $V_{oc}$ variation with temperature.

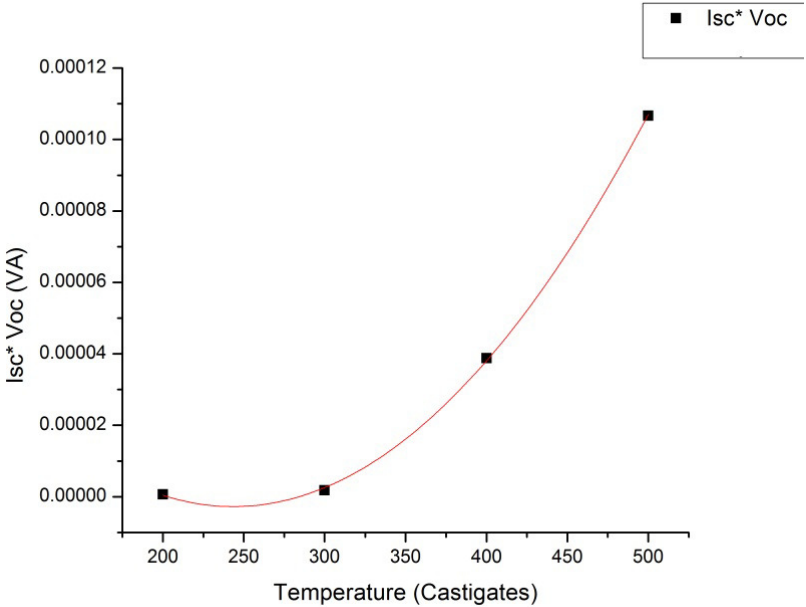

Figure 4: $I_{sc} * V_{oc}$ Variation With temperature.

According to figure 4, the output power ($I_{sc} * V_{oc}$) increases with the annealing temperature. This occurs due to the increase of the nano particle size with annealing temperature; these particles can absorb lower wavelength of the electromagnetic spectrum than the nano particles with lover radius. Therefore, for solar cell applications, the CdS samples annealed at higher temperatures provide better efficiencies.

Due to the following reactions taking place at CdS thin film and counter electrode in $KI/I_2$ electrolyte, the charge carriers flow through to circuit. These charge carriers originate the photocurrent and photovoltage.

At counter electrode,

$I_3^- + 2e \rightarrow 3I^-$

At n-type semiconductor (CdS film)

$I_2 + 2e \rightarrow 2I^-$

## 4. Conclusion:

The CdS bond formation and the output power of our thin film samples increase with the annealing temperature. The efficiency may increase due to the effect of increase of CdS bond formation and improvement of particle size with the annealing temperature. Due to the high band gap of CdS, the photovoltage is higher, and the photocurrent is lower. Although the photovoltage of our CdS thin films samples is in the volt range, the photocurrent is in the microampere range. As a result, the output power of the photocell with CdS is low. This same phenomenon was observed by us for reactive dc sputtered ZnO films with high band gap [15]. Redox couples in $KI/I_2$ electrolyte provide the charge transformation in the liquid junction photocell.